\documentstyle[mprocl]{article}

\def\Journal#1#2#3#4{{#1} {\bf #2}, #3 (#4)}

\def\be{\begin{equation}}
\def\ee{\end{equation}}
\def\bea{\begin{eqnarray}}
\def\eea{\end{eqnarray}}

\begin{document}

\title{The Structure of Singularity in Gravitational Collapse}

\author{ S. Jhingan }

\address{Theoretical Astrophysics Group,\\ Tata Institute of Fundamental 
Research, \\ Colaba, Mumbai 400 005, INDIA}

\maketitle\abstracts{We will describe here the structure of singularity 
forming in gravitational collapse of spherically symmetric inhomogeneous dust.
Such a collapse is described by the Tolman-Bondi-Lema{\^i}tre metric.
The main new result here relates, in a general way, the formation of black 
holes and naked shell-focusing singularities resulting as the final fate of 
such a collapse to the generic form of regular initial data. Such a data is
characterized in terms of the density and velocity profiles of the matter, 
specified on an initial time slice from which the collapse commences.
We show that given any generic density profile at the initial time slice, 
there exists a corresponding velocity profile which gives rise to a strong
curvature naked singularity. This establishes that strong naked singularities
arise for a generic density profile. We also establish here that similar 
results hold for black hole formation as well. Keeping the model to be
spherically symmetric  we also consider more general form of matter fields,
i.e. equation of state $p = k\rho$. We will analyse here the nature of 
non-central singularity forming due to collapse of spherically symmetric
perfect fluid subject to weak energy condition.}

\section{Introduction}

The gravitational collapse of a spherically symmetric 
homogeneous dust can result into a black hole in the space-time\cite{OSD},
characterized by the presence of an event horizon, and the space-time 
singularity at the center, which is covered by the horizon. 
This model provides the basic motivation for the black hole physics, and the 
cosmic censorship conjecture \cite{Pen}, which states that even when the 
assumptions contained in the above case are relaxed, in the form of either 
perturbations in the symmetry or form of matter etc., the outcome would still 
be a black hole in generic situations. It is clear that the assumption of 
homogeneity is only an idealization, and realistic density profiles, 
for massive objects such as stars will have inhomogeneous density distribution
, peaked typically at the center of the object. In section 2 we analyse 
inhomogeneous dust models, called Tolman-Bondi-Lema{\^i}tre 
({\bf TBL})\cite{Tolman} models.

\noindent{The use of a pressureless fluid in all these models is an important 
assumption. Analysing the Einstein equations for a more general equation of 
state, subject to suitable energy conditions, is astrophysically more 
important. Using analytical methods, similar to  those developed by Podurets
\cite{Podure}, we demonstrate the role of  pressure in determining
whether a naked singularity or a black hole will form in the collapse.}

\section{Tolman-Bondi-Lema{\^i}tre ({\bf TBL}) models}

The TBL space-times are spherically symmetric manifolds
($\cal M$,g), with metric of the form,
\be
ds^2 = -dt^2 + \frac{R'^2}{1 + f(r)} dr^2 + R^2 d\Omega^2,
\ee
and energy-momentum tensor of the form of a perfect fluid with
equation of state p=0, given by
\be
{T^{ij} = \epsilon \,\delta^{i}_{t}\delta^{j}_{t}}.
\ee
Here $\epsilon$ and R are functions of  $r$ and  $t$, and
$d\Omega^2$ is the metric on the 2-sphere. The Einstein
equations become 
\be
{\dot R}^2 = \frac{F(r)}{R}  + f(r), \; \; \epsilon(r,t)= \frac{F'(r)}{R^2 R'},
\ee
where the dot and prime signify partial derivatives with respect to
$t$ and $r$ respectively. It is seen that the energy density blows up either 
at the ``shell-focusing singularity'' $R=0$, or when $R'=0$ which corresponds 
to a ``shell-crossing singularity'' in the space-time.
Equation (3) leads to the interpretation of $F(r)$ and $f(r)$ as mass and 
energy function respectively. The model is said to be bound,
unbound, or marginally bound if $f(r)$ is less than zero, greater than
zero, or equal to zero.

\noindent{The integrated form of equation (3)  is given by   
\be
t - t_s(r) = -\frac{R^{3/2} G(-fR/F)}{\sqrt F}
\ee
where $G(y)$  is a real, positive and smooth function which is bounded,
monotonically increasing and strictly convex. $t_s(r)$ is a constant of 
integration, fixed by the choice of scaling on the initial surface
i.e. $ R(0,r)= r$. The time $t=t_s(r)$ corresponds to the value $R=0$
where the area of the matter shell at a constant value of the coordinate $r$ 
vanishes, which corresponds to the physical space-time singularity. Thus the 
range of coordinates is given by
\be
0 \, \leq \, r \, < \, r_c, \; \; \;  -\infty \, \leq \, t \, < \,
t_{s}(r),
\ee
where $r=r_c$ denotes the boundary of the dust cloud where the
solution is matched to the exterior Schwarzchild solution.}

\noindent{It can be shown that\cite{JosJhi} for a generic expandable density 
and velocity profile, on the initial regular hypersurface, of the form
\be
\rho(r) = \sum_{n=0}^{\infty} \rho_n r^n , \; \; i.e. \; \; F(r) = 
\sum_{n=0}^{\infty}F_n r^{n+3},\; \; and \; \; f(r) = 
\sum_{n=2}^{\infty}f_nr^{n}.
\ee
the existence of locally naked singularity (black hole) is related to 
existence (absence) of a real positive root to the  quartic equation
\be
(\alpha - 1) x^4 + \sqrt{\Lambda_0} x^3 -\Theta_0 x + \sqrt{\Lambda_0}
\Theta_0 = 0
\ee
where  $x=\sqrt{(R/r^{\alpha})}$, $\Lambda = \frac{F}{r^{\alpha}}$ and $\Theta 
\equiv \frac{t'_s \sqrt \Lambda}{r^{\alpha-1}}$. Subscript ``0'' denotes values
of the functions specified on initial regular hypersurface at $r=0$. Alpha is 
a free parameter introduced to analyse the characteristic curves and can be 
uniquely fixed for a given initial data. Also it 
is shown\cite{JosJhi} that both naked singularity and black hole can develop
from generic form of initial data.}

\section{Perfect fluid models}

We study here the formation of singularities in the perfect fluid
space-time, equation of state $p = k \rho$, subject to weak energy condition.
The metric in comoving coordinates is of the form
\be
ds^{2} = (\rho^{\frac{-2k}{1+k}}) dt^{2} - \frac{(\rho^{\frac{-2}{1+k}})}{R^4}
 dr^{2} - R^{2} 
d\Omega^{2} 
\ee
where $\rho (r,t)$ denotes energy density and $R(r,t)$ is area coordinate.
Einstein's equations assumes the form
\be
\begin{array}{rcl}
m & = & \frac{R}{2}(1 + \rho^{\frac{2k}{1+k}}(\dot{R})^2 - R^4 
(\rho^{\frac{2}{1+k}})(R')^2) \\
& & \\
\dot{m} & = & -4\pi p_{r}R^{2}\dot{R} \\
& & \\
m' & = & 4\pi \rho R^{2} R' \\
& & \\
\end{array}
\ee
Weak energy condition i.e. $\rho \geq 0$; $\rho + p_r \geq 0$ and $\rho + 
p_{\theta} \geq 0$, implies $k \geq -1$. 
This allows for the possibility of negative pressure in extreme conditions
during gravitational collapse. $\dot{m}$ equation above indicates that for 
collapse models ($\dot{R} < 0$) positivity of energy density implies that 
negative values of $k$ can make ``m'' vanish in the limit of approach to 
singularity. It can be shown that\cite{Coop} we can have
a approximate solution near singularity and  
for non-central points on the singularity curve mass function ``m'' 
necessarily vanishes, as we approach singularity, for
$-1 \leq k \leq -1/3$. Also the singularity is always naked in this range
independent of rest of the initial data. 

\noindent{This indicates the possibility that 
the earlier results for the spherical gravitational collapse for dust equation
of state might generalize to the case of perfect-fluid space-times.}

\section*{Acknowledgments}
I thank P. S. Joshi, T. P. Singh and F. Cooperstock for useful discussions and
acknowledge the financial support by TIFR and UNESCO grant for participating
in Marcel Grossmann meeting.

\end{document}